\documentclass[journal,comsoc]{IEEEtran}
\usepackage[T1]{fontenc}
\usepackage{algorithmic}
\usepackage[titlenumbered,ruled]{algorithm2e}
\usepackage{caption}
\usepackage[flushleft]{threeparttable}
\usepackage{cite}
\usepackage{cleveref}

\ifCLASSINFOpdf
\else
\fi
\usepackage{amsmath}
\usepackage[cmintegrals]{newtxmath}
\usepackage{amsmath}
\usepackage{array}
\usepackage{balance}
\usepackage{color}
\usepackage{graphicx, floatrow}
\usepackage{subfig}
\usepackage{booktabs}
\usepackage{multirow}
\usepackage{bm}
\usepackage[hyphens]{url}
\usepackage{float}
\floatstyle{plaintop}
\restylefloat{table}
\usepackage{etoolbox}
\makeatletter
\patchcmd{\@makecaption}
{\scshape}
{}
{}
{}
\makeatother
\usepackage{caption}
\begin{document}
\title{Optimizing Space-Air-Ground Integrated Networks by Artificial Intelligence}
\author{Nei Kato,~\IEEEmembership{Fellow,~IEEE},
Zubair Md. Fadlullah,~\IEEEmembership{Senior Member,~IEEE,}
Fengxiao Tang,~\IEEEmembership{Student Member,~IEEE,}
Bomin Mao,~\IEEEmembership{Student Member,~IEEE,}
Shigenori Tani,~\IEEEmembership{Member,~IEEE,}
Atsushi Okamura,~\IEEEmembership{Senior Member,~IEEE,}
Jiajia Liu,~\IEEEmembership{Senior Member,~IEEE}

\thanks{Nei Kato, Zubair Md. Fadlullah, Fengxiao Tang, and Bomin Mao are with the Graduate School of Information Sciences, Tohoku University, Sendai, Japan. 
Emails: 
			\{kato, zubair, fengxiao.tang, and bomin.mao\}@it.is.tohoku.ac.jp}
\thanks{Shigenori Tani and Atsushi Okamura are with Information Technology R\&D center, Mitsubishi Electric Corporation, Kamakura, Japan. Emails: \{Tani.Shigenori@eb, Okamura.Atsushi@bc\}.MitsubishiElectric.co.jp}

\thanks{Jiajia Liu is with the State Key Laboratory of Integrated Services Networks, School of Cyber Engineering, Xidian University, Xi'an 710071, China.
Email: liujiajia@xidian.edu.cn}}

\markboth{IEEE Wireless Communications,~Vol.~XX, No.~XX, July~2018}%
{Shell \MakeLowercase{\textit{et al.}}: Bare Demo of IEEEtran.cls for IEEE Communications Society Journals}

\maketitle

\begin{abstract}
It is widely acknowledged that the development of traditional terrestrial communication technologies cannot provide all users with fair and high quality services due to the scarce network resource and limited coverage areas. To complement the terrestrial connection, especially for users in rural, disaster-stricken, or other difficult-to-serve areas, satellites, unmanned aerial vehicles (UAVs), and balloons have been utilized to relay the communication signals. On the basis, Space-Air-Ground Integrated Networks (SAGINs) have been proposed to improve the users' Quality of Experience (QoE). However, compared with existing networks such as ad hoc networks and cellular networks, the SAGINs are much more complex due to the various characteristics of three network segments. To improve the performance of SAGINs, researchers are facing many unprecedented challenges. In this paper, we propose the Artificial Intelligence (AI) technique to optimize the SAGINs, as the AI technique has shown its predominant advantages in many applications. We first analyze several main challenges of SAGINs and explain how these problems can be solved by AI. Then, we consider the satellite traffic balance as an example and propose a deep learning based method to improve the traffic control performance. Simulation results evaluate that the deep learning technique can be an efficient tool to improve the performance of SAGINs.

\end{abstract}


\IEEEpeerreviewmaketitle

\section{Introduction}
\label{intro}
\begin{figure*}[!t]
	\centering
	\includegraphics[scale=0.5]{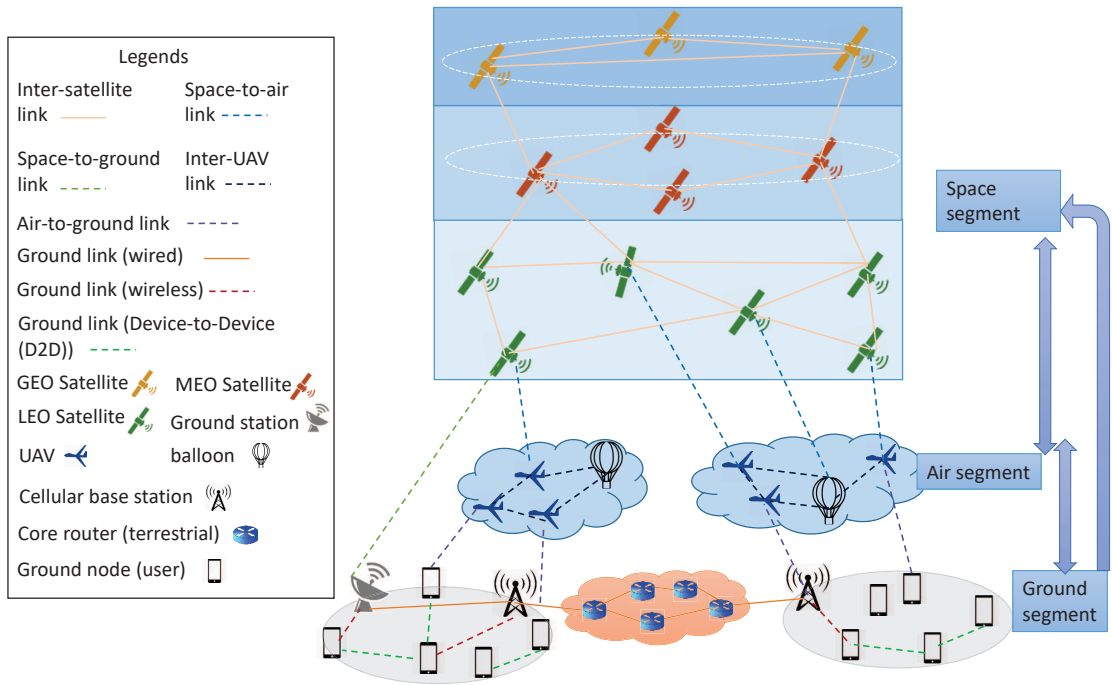}
	\caption{The architecture of the Satellite-Air-Ground Integrated Network.}
	\label{sagin}
\end{figure*}

\IEEEPARstart{I}{n} recent years, significant achievements have been achieved in terrestrial communication systems to offer users Internet access with much higher data rate and lower latency~\cite{shen-5g}. The next generation wireless system, 5G, is announced 20 times faster than 4G with ultra-densely deployed base stations. However, this does not mean that all users can enjoy such kind of high quality Internet services at any time due to the limited network capacities and coverage areas. For instance, in rural areas which lack the terrestrial high-speed broadband infrastructure because of the high expense, users still suffer the narrow-band network access~\cite{satellite-survey}. Moreover, users in a stadium or participating concerts cannot enjoy stable Internet service. To address these problems, industry have made many endeavors, such as the Google's Project Loon and Facebook's Aquila, which provide Internet access through balloons and drones located in the air, respectively~\cite{air-com}. As shown in Fig.~\ref{sagin}, compared with the ground segment, the air and space communication systems have much larger coverage, which have been utilized to provide Internet service to islands, isolated mountainous areas, as well as the disaster areas. 

However, different from the ground networks, the air networks and space networks have their own shortcomings. Users of satellite communication systems have to tolerate the long propagation latency, while the air networks have limited capacity and unstable links. Moreover, both the two networks are of high mobility. On the other hand, the terrestrial networks have the highest throughput and most resource. As the three networks at different altitudes can complement each other, researchers proposed the integrated network termed Space-Air-Ground Integrated Network (SAGIN) as shown in Fig.~\ref{sagin} to provide users with improved and flexible end to end services~\cite{liu-sagin}. It can be found that the SAGIN is a hierarchical network, with the satellites at the top, the unmanned aerial vehicles (UAVs), balloons, and airships in the air, and the ground segment. Notice that the space segment consists of the satellites, constellations, as well as the corresponding terrestrial infrastructures, such as the ground stations and the control centers.  

Aided by the air and space segments, the SAGIN in Fig.\ref{sagin} enables the packets generated by the ground nodes to be transmitted to destinations via multiple paths of various quality. Therefore, the three heterogeneous networks can offer differential packet transmissions to meet various service requirements, which is extremely important for the Internet of Things (IoT) network. Moreover, the multi-layered satellite communication systems which consists of the Geostationary Earth Orbit (GEO) satellites, the Medium Earth Orbit (MEO) satellites, and the Low Earth Orbit (LEO) satellites, can utilize the multi-cast and broadcast techniques to improve the network capacity, which can significantly alleviate the growing traffic burden~\cite{liu-sagin,satellite-survey}. However, the network design and optimization in SAGIN are facing lots of challenges due to the inherent characteristics of heterogeneity, self-organization, and time-variability~\cite{liu-sagin}. To improve the end to end Quality of Experience (QoE), it would happen frequently that researchers have to conduct the analysis from all three parts since many transmission in SAGINs are collaboratively conducted by the base stations in the ground, the UAVs in the air, and the satellites in the space. In such case, more factors have to be considered~\cite{liu-sagin} compared with the same problem in conventional ground communication systems.

Since the SAGIN is more complex than the current terrestrial networks, we need to consider more efficient techniques to optimize the SAGIN performance. The Artificial Intelligence (AI) technique has always been a hot topic due to its predominant performance in both industry and academia. In recent years, the rapid development of deep learning refreshes human beings' realization of AI technique. Google's Alphago and Alphago Zero force humans to realize the fact that current AI technology can not only master the board game go even though the complex game is concerned with $2\times{10^{170}}$ legal positions, but also beat human beings~\cite{zubair-survey}. Besides the excellent performance in games, deep learning has also shown overwhelming advantages over human beings in various industrial applications, such as the disease diagnose, automatic drive, and industrial control. Moreover, the deep learning technique has also been utilized in satellites for many applications, such as the image processing. Furthermore, researchers have conducted lots of research on the utilization of deep learning to improve the network performance~\cite{wcm1,zubair-survey}. The deep learning technique is demonstrated to outperform conventional methods. It should also be noted that current hardware development has also enabled the deep learning technique to run successfully on communication infrastructures~\cite{mao2017routing}. Considering all these factors and examples, it is a perspective direction to adopt the deep learning technique to optimize the network performance of SAGIN.

In this paper, we consider the deep learning technique to optimize the performance of SAGIN. We first analyze several urgent problems and challenges which are most related to the SAGIN performance. Then, we introduce some existing research on the applications of deep learning. After that, we give an example to illustrate how to adopt the deep learning technique to choose paths for the satellite networks. We then discuss the future directions about the application of deep learning in SAGINs. Finally, we conclude the whole paper.

\begin{figure*}[!t]
	\centering
	\includegraphics[scale=0.4]{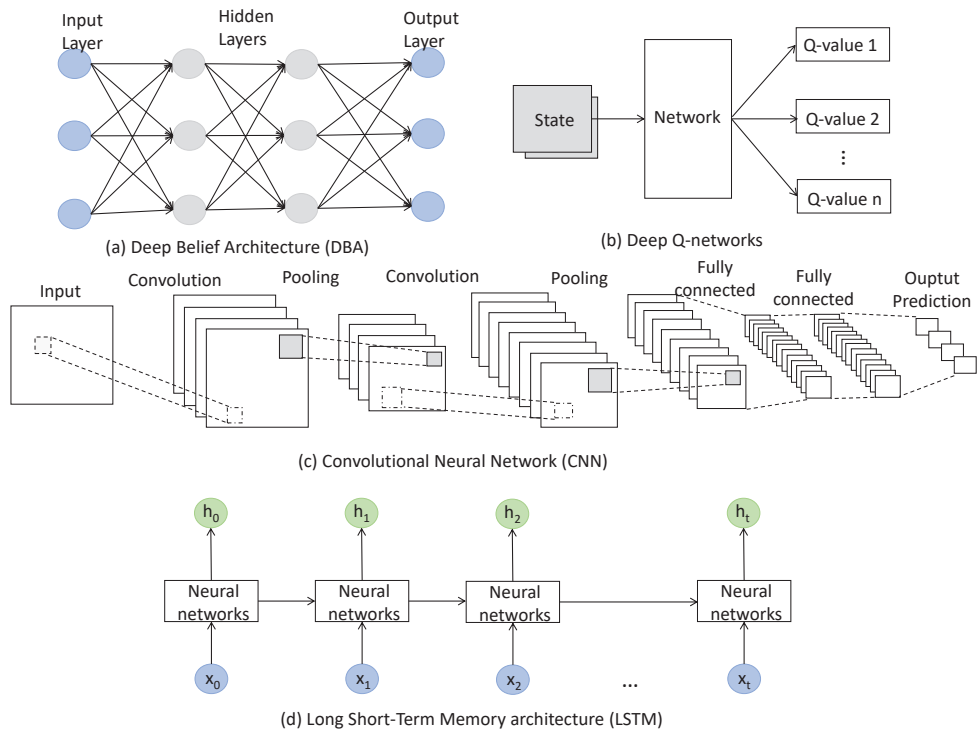}
	\caption{The commonly utilized deep learning architectures.}
	\label{dl-architecture}
\end{figure*}

\section{Challenges}
\label{challenges}

It has been widely acknowledged that current ground networks are facing exponentially increasing traffic generated by various things connected to the network~\cite{mao2017routing}. The advent of SAGIN can significantly alleviate the challenge and provide users with more choices. Since recent developments in the industry such as SpaceX and the next generation of GEO fixed satellite systems, have considerably decreased the communication cost and increased the available capacity~\cite{liu-sagin}, it can be imagined that the future earth will be surrounded by large volumes of satellites. Due to the three dimensions and the high mobility of space segment as well as the air segment, the SAGIN is much more complex than current ground networks. In this section, we discuss some challenges which is deeply related to the SAGIN performance.

\textbf{\emph{Network Control:}} The control manner of the SAGIN is one of the most important factors which affect the network performance directly. The distributed management can reduce the possibility of bottleneck and response time, while the cooperative operation among devices increases the network complexity. On the other hand, the centralized control manner can simplify the network structure, whilst the response delay significantly affects the network performance. It is reasonable to assume that both two control manners will be cooperatively utilized in future SAGIN to increase the network tolerance to different mistakes. Then, the different control manners lead to the diversity in other network management algorithms, which further complicates the network management. Moreover, since the SAGIN consists of heterogenous networks which are composed various infrastructure, the network integration significantly affects the QoE. Due to the inherent heterogeneity as well as the high mobility, the management of SAGIN is exposed to extreme difficulties~\cite{liu-sagin}.

\begin{table*}[t]
	\centering
	\caption{Some existing networking research based on deep learning}
	\label{dl-application}
\begin{tabular}{|l|l|l|l|l|l|}
	\hline
	Purpose                                                                        & Strategy                                                     & \begin{tabular}[c]{@{}l@{}}Deep learning\\ architecture\end{tabular} & \begin{tabular}[c]{@{}l@{}}Training\\ manner\end{tabular}           & \begin{tabular}[c]{@{}l@{}}Training\\ time\end{tabular} & Challenges for SAGIN                                                                                                                                                                                                                                                                                                                                                                                                                                                   \\ \hline
	\begin{tabular}[c]{@{}l@{}}Traffic \\ control\end{tabular}                     & Routing                                                      & DBA                                                                  & \begin{tabular}[c]{@{}l@{}}Supervised\\ learning\end{tabular}       & Offline                                                 & \multirow{4}{*}{\begin{tabular}[c]{@{}l@{}}Each individual segment has its inherent\\ requirements and tailored techniques and\\ protocols to cater to those requirements.\\ Hence, the deep learning based conventional\\ applications for the individual segments\\ cannot provide a smart, holistic solution. \\ Therefore, in this article, we aim to investigate\\ how to leverage deep learning for the integrated\\ space, air, and gound networks.\end{tabular}} \\ \cline{1-5}
	\multirow{2}{*}{\begin{tabular}[c]{@{}l@{}}Resource\\ allocation\end{tabular}} & \begin{tabular}[c]{@{}l@{}}Cache\\ allocation\end{tabular}   & \begin{tabular}[c]{@{}l@{}}Deep\\ Q-network\end{tabular}             & \begin{tabular}[c]{@{}l@{}}Reinforcement\\ \\ learning\end{tabular} & \multirow{2}{*}{Offline}                                &                                                                                                                                                                                                                                                                                                                                                                                                                                                                          \\ \cline{2-4}
	& \begin{tabular}[c]{@{}l@{}}Channel\\ assignment\end{tabular} & \begin{tabular}[c]{@{}l@{}}DBA,\\ CNN\end{tabular}                   & \begin{tabular}[c]{@{}l@{}}Supervised\\ \\ learning\end{tabular}    &                                                         &                                                                                                                                                                                                                                                                                                                                                                                                                                                                          \\ \cline{1-5}
	Security                                                                       & \begin{tabular}[c]{@{}l@{}}Anomoly \\ detection\end{tabular}    & LSTM                                                                 & \begin{tabular}[c]{@{}l@{}}Unsupervised \\ learning\end{tabular}    & Online                                                  &                                                                                                                                                                                                                                                                                                                                                                                                                                                                          \\ \hline
\end{tabular}
\end{table*}

\textbf{\emph{Spectrum Management:}} As we know, the quality of wireless communication is affected by the propagation medium. Compared with terrestrial networks, the propagation medium in SAGIN is much more diverse since it is highly variant in time, frequency, and space. Due to the different impacts of the propagation medium as well as transmission requirements, multiple frequency bands are cooperatively utilized in SAGIN to improve the network performance. Since the propagation medium of SAGIN is distinctly different from the well studied terrestrial communication systems and the high variance leads to the frequent changes, it still needs more endeavors to manage the spectrum in the SAGIN~\cite{liu-sagin}. 
Moreover, the channel resource allocation is still one of the most important factors to affect the network performance despite of the abundant research on this topic for ground networks. As mentioned above, the earth will be surrounded by massive satellites as well as aerial infrastructure in large volume. As the frequency bands are already congested, Internet Service Providers (ISPs) have to consider sharing the same frequency bands for many different types of communications. For instance, L-band is extensively utilized for space-ground, space-air, and air-ground communications. Then, the mutual interference caused by different communication process utilizing the same channels or adjacent channels should be considered when optimizing the SAGIN performance~\cite{l-band}. However, the inherent heterogeneity and high mobility of SAGIN make the problem more complex, meaning that more efficient techniques should be considered compared with that of terrestrial networks. 

\textbf{\emph{Energy Management:}} Different from some devices in the ground which can be charged or connected to the power at any time, the aerial and space infrastructure as well as some IoT sensors cannot be connected to a power station. These devices are powered by the solar or battery. For instance, current satellites have solar cells to absorb and convert the solar energy, and then utilize batteries to save the energy for usage when moving to the shadow area. To improve the energy efficiency can increase the usage time of the satellites since the total times of complete charge of battery cells are limited. 
However, different from the ground segment which usually just improves communication mechanisms to save the energy, the totally different propagation medium, the intense radiation, and space-variant temperature also affect the energy consumptions of satellite segments~\cite{energy-routing}. For the aerial infrastructure, optimizing the energy efficiency is more critical than conventional terrestrial communication system. Moreover, while the power consumption in terrestrial communication systems are mainly related to the communication related functions, the majority power consumption for the aerial segments are to maintain the aloft and support their mobility~\cite{EE-UAV}. Therefore, the energy management is a critical but challenging topic for SAGIN.

\textbf{\emph{Routing and Handover Management:}} Since the SAGIN is a multi-layer network, there exists multiple paths for most source and destination pairs. On the one hand, the multiple paths can be utilized for different types of traffic transmission for meeting the diverse service requirements. On the other hand, this increases the difficulty in routing strategy design since the network researchers need to consider the service requirements and the performance of paths in terms of packet loss rate, end-to-end delay, and throughput~\cite{uav-survey}. Another challenge for routing design is the high mobility of SAGIN which leads to uncertainties in locations of mobile devices. Moreover, the high mobility also results in the frequent handover, which is a challenge for ensuring the seamless transmissions. As the SAGIN consists of heterogeneous components and different segments have various coverage areas, different handover schemes should be proposed with the properties of corresponding segments considered. For instance, the handover in UAV-based communication systems should consider the work states of UAVs since they are often shutdown to save power. Furthermore, efficient prediction algorithms are also required due to the location uncertainties~\cite{liu-sagin}.  

\textbf{\emph{Security Guarantee:}} As a cooperative network with open links and dynamic topologies, SAGIN are facing many security threatens. Even though the SAGIN is required to offer reliable and secure communications due to the large amount of sensitive data, it is still difficult to provide high security level, for which the main reasons are multi-fold. First, the conventional communication protocols, e.g. TCP/IP, have been revised to integrate diverse technologies for performance improvement, which leads to new problems in IP security mechanism. Second, the encryption operations often results in more processing delay and transmission delay, which may hamper the real-time communications. And the higher the security level, the more time it costs. Third, the high mobility of the SAGIN produces many challenges to the network security. For instance, the frequent handover makes the secure routing more difficult to be realized. And the SAGIN becomes more vulnerable to jamming, which is difficult to be addressed because of the large coverage area~\cite{liu-sagin}.

\section{Related Research on AI based Networking Optimizations}
\label{ai-applications}

Similar to the SAGIN, current terrestrial networks also have the above-mentioned but less complex problems. To address these problems, the deep learning technique has been a perspective direction to optimize the network performance in recent years. And the results have evaluated the advantages of the deep learning over traditional strategies. In this section, we make some analysis about the existing research and then discuss the necessary modifications for adoption in the SAGIN. It should be noted that even though existing proposals cannot be applied in SAGIN directly, they can provide a solid foundation for future research.

\subsection{Intelligent Traffic Control}
\label{routing-ai}
As the network traffic is exponentially increasing, traffic control is a hot research topic. Our previous work~\cite{wcm1} proposed a deep learning based routing strategy which utilizes the Deep Belief Architecture (DBA) as shown in Fig.~\ref{dl-architecture}(a) to predict the paths. In the proposal, the input and output of the DBA are the traffic patterns and the next node, respectively. The simulation results show that the proposal can significantly improve the network throughput and reduce the average delay per hop. To enable the deployment of the proposed deep learning based routing strategy, in~\cite{mao2017routing}, we analyzed the Graphics Processing Unit (GPU) accelerated Software Defined Routers (SDRs). Theoretical analysis shows that time cost for training and running the strategy with GPU is much lower than that with CPU. Moreover, the SDR architecture also provides more opportunities to improve the flexibility. Besides the next node prediction, the deep learning technique is also utilized for the traffic forecast, which can be adopted for routing design and following resource allocation ~\cite{tang-poc}.

\subsection{Intelligent Resource Allocation}
\label{channel-assignment-ai}
Resource allocation is an important factor which directly impacts on the network performance. In current terrestrial networks, the network resource allocation is usually concerned with two types of resource: the communication resource including channels and bandwidth, and the computation resource which consists of the processing power and memory. The authors of~\cite{8290944} discussed the balance between two conflicting factors: the network cost and users' mean opinion score (MOS) to improve the network QoE. In this paper, authors utilized the deep reinforcement learning technique to change the cache location of content. Specifically, the network states which consists of the transmission rates and cache condition are utilized as the input of the deep Q-networks shown in Fig.~\ref{dl-architecture}(b). Then, the agent can choose the best action according to the output Q-values of different actions.  Simulation results demonstrate the agent can find the best decision to optimize the proposed reward function after trial and error. As we mentioned above, the traffic prediction is usually conducted for many network management tasks. Our previous work~\cite{tang-poc} proposed a deep learning based partially overlapped channel assignment strategy for the IoT network. The proposal consists of two steps: utilize the DBA as shown in Fig.~\ref{dl-architecture}(a) to predict the network traffic in next time interval, then adopt the Convolutional Neural Network (CNN) shown in Fig.~\ref{dl-architecture}(c) to choose the suitable channels according to the predicted traffic. Since the input of the DBA considers the properties of the IoT traffic, the prediction accuracy is much higher than conventional methods, which leads to the performance improvement of channel assignment.

\begin{figure}[!t]
	\centering
	\includegraphics[scale=0.35]{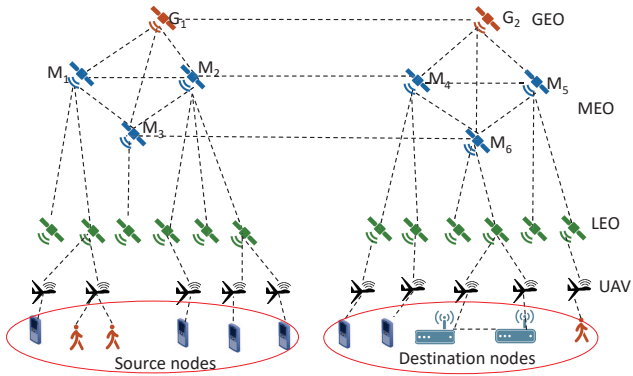}
	\caption{The utilized network topology in the simulation.}
	\label{simu-topo}
\end{figure}

\subsection{Smart Anomaly Detection}
\label{security-ai}
Since future IoT network will be abounded with a large volume of users' data, it is critically important to increase the security and privacy level of corresponding network services. In existing ground networks, deep learning has also been studied for improving the network security level. In ~\cite{DBLP:journals/corr/abs-1710-00811}, the authors proposed a deep learning based approach to detect anomalous network activity. In the paper, the deep neural network (DNN) is adopted to extract the features of users' activities from the system logs. Then, the feature vectors are input to the long short-term memory (LSTM) as shown in Fig.~\ref{dl-architecture}(d) to measure the anomaly score. Since the users' activities are often unpredictable over seconds to hours, authors in this paper utilized an online unsupervised training fashion to train the deep learning architectures, by which the models can adapt to the changing patterns in the data. Simulation results show that the proposed strategy has very high accuracy rate and can significantly reduce the analyst workloads.

After discussing the deep learning based strategies to solve various problems in current terrestrial networks, it can be clearly found that the deep learning technique can be utilized to optimize the network performance. As shown in Table.~\ref{dl-application}, we can find that the various deep learning architectures as well as the different training manners enable this technique to be flexibly adopted in heterogeneous scenarios. On the other hand, current research is all concentrated on the terrestrial networks and does not consider the characteristics of future SAGIN. The heterogeneity of different segments in SAGIN is an important factor to impact on the network performance which also leads to more complexity. The inherent requirements of each individual segment as well as the tailored techniques need to be considered when desinging the deep learning based algorithms for SAGIN.

\section{Case Study on Deep Learning Aided SAGIN}
\label{example}

In this section, we take the routing problem as an example to show how the deep learning can be utilized to optimized the SAGIN performance. Due to the huge scale of SAGIN as well as some uncertainties of future aerial and space segments' structures, in this paper, we just use a simplified network structure shown in Fig.~\ref{simu-topo} as the considered scenario. In the figure, we can find the hierarchical structure consists of 5 layers. In the space segment, there are 2 GEOs, 6 MEOs, and 12 LEOs, respectively. The aerial segment is composed of 120 UAVs which get the most applications. Since researchers have spent most energy in studying the ground segment, the terrestrial communication technologies are relatively mature. In this paper, we simplify the ground segments and only consider the connections between the ground nodes and the UAVs. And if the ground nodes need to transmit the packets via the space segments, they will first upload the packets to the UAVs. The total number of ground nodes are 3,200 and they are evenly distributed. Due to the inherent heterogeneity of the three segments, it is reasonable to consider that the routing design follows different strategies in different segments. As the deep learning technique has been studied to improve the terrestrial and air segments, in this paper, we focus on the performance optimization of the satellite network with deep learning. Since the simulation platform is a workstation with Intel Core i7-6900K CPU, 64GB Random Access Memory (RAM), and Nvidia Geforce Titan X GPU, it is reasonable to just consider the inter-layer links in the layers of GEO and MEO as well as the links connecting adjacent layers, which can still demonstrate the advantages of our proposed proof-of-concept. The values of main parameters in the considered topology are given in Table~\ref{simu-para}.

\begin{table}
	\centering
	\caption{The values of main parameters utilized in the simulation}
	\label{simu-para}
\begin{tabular}{|c|c|c|}
	\hline
	\multirow{2}{*}{\begin{tabular}[c]{@{}c@{}}GEO/\\ MEO\end{tabular}} & \begin{tabular}[c]{@{}c@{}}Frequency band of \\ connected links\end{tabular} & Ka band \\ \cline{2-3} 
	& Channel capacity                                                             & 1Gbps   \\ \hline
	\multirow{2}{*}{LEO}                                                & Downlink                                                                     & L band  \\ \cline{2-3} 
	& Channel capacity                                                             & 120Mbps \\ \hline
	\multirow{2}{*}{CNN}                                                & \#Convolutional layers                                                       & 3       \\ \cline{2-3} 
	& \#Fully connected layers                                                     & 3       \\ \hline
\end{tabular}
\end{table}

\begin{figure*}[!ht]
	\centering 
	\subfloat[Comparison of network throughput for conventional routing strategy and our proposal.]{ \label{per1}	\includegraphics[width=7cm]{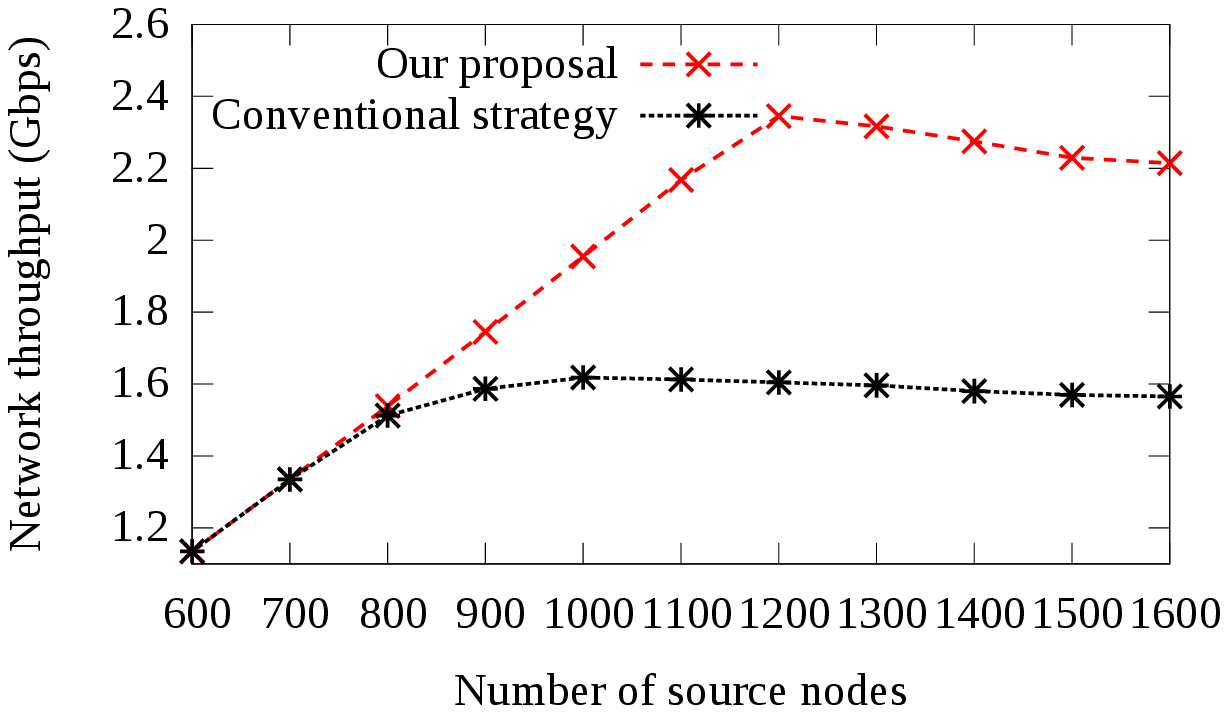}} \hspace{1.5ex}
	\subfloat[Comparison of network packet loss rate for conventional routing strategy and our proposal.]{ \label{per2}	\includegraphics[width=7cm]{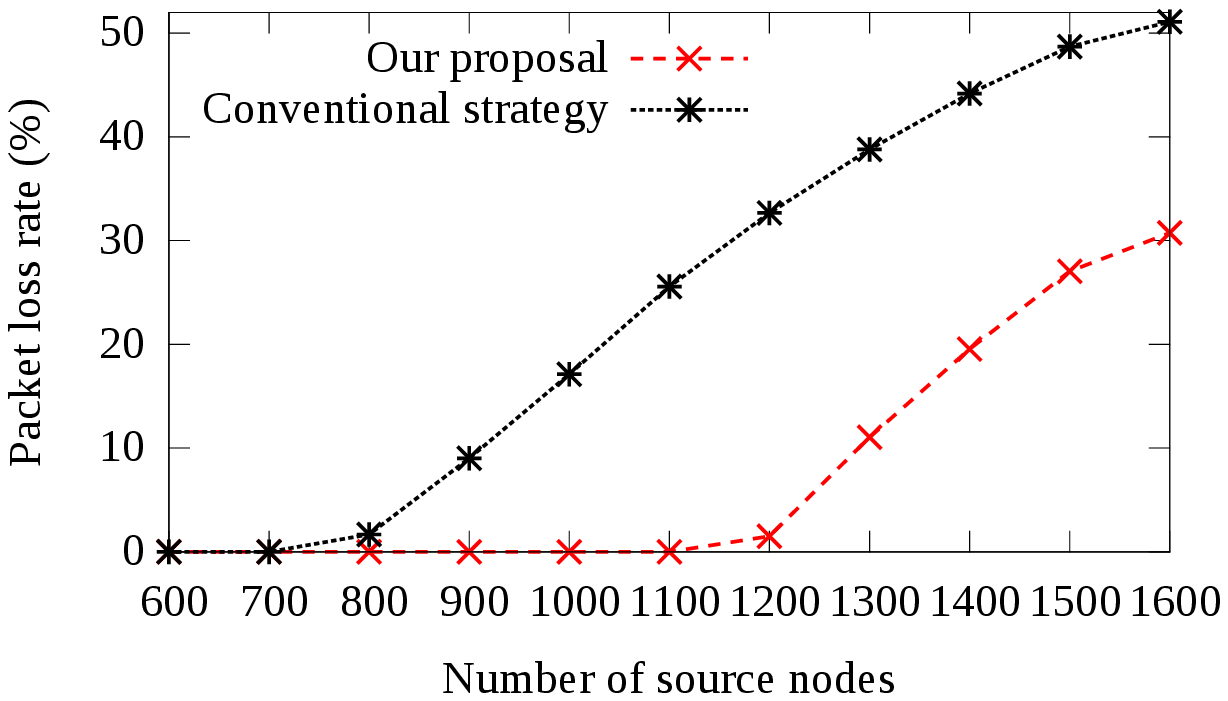}}		
	\caption{The performance of our proposal in terms of network throughput and packet loss rate compared with the conventional routing strategy.}
	\label{fig-result}
\end{figure*}

In Fig.~\ref{simu-topo}, the network can be separated into two parts: the left part and the right part covered by G1 and G2, respectively. Even though there are lots of simultaneous communication process in the SAGIN, we only focus on the packets generated by the ground nodes in the left part and destined for the ground nodes in the right part via the satellite networks. Therefore, the ground nodes are evenly separated into two groups: 1,600 source nodes in the left part and 1,600 destination nodes in the right part. We consider various number of source nodes which need to transmit packets to the destination nodes to change the traffic load of the space segment. For each source node communicating with the destination nodes, its packet generation rate is 2Mbps. According to Fig.~\ref{simu-topo}, the inter-layer connections in the MEO and GEO satellites enable these packets to be transmitted to the right part via multiple paths. For instance, the packets in $\mathit{M_1}$ destined for $\mathit{M_5}$ can be transmitted via $\mathit{M_2}$ and $\mathit{M_4}$, or via $\mathit{G_1}$ and $\mathit{G_2}$. As the long propagation delay caused by the distances among satellites deteriorates the performance seriously, the physical distance is one of the most important metrics impacting on the routing design. Therefore, according to the conventional shortest path (SP) algorithm, the path via the MEO satellites usually has higher priority over that through the GEO satellites due to the less distance. However, this actually does not lead to better performance, especially when $\mathit{M_1}$, $\mathit{M_2}$, and $\mathit{M_3}$ all have heavy traffic towards the right section. If we consider the periodical bursty traffic as shown in~\cite{wcm2}, the MEO satellites connecting two parts, $\mathit{M_2}$, $\mathit{M_3}$, $\mathit{M_4}$, and $\mathit{M_6}$, will be congested frequently. On the other hand, if the satellites can learn the rules according to the traffic patterns, the MEO satellites can transfer some traffic to GEO satellites for the traffic balance. Here, we consider the deep learning technique to reach this goal.

We consider the origin-destination (OD) pair which consists of a MEO satellite in the left part and right part denoting the origin and destination, respectively. Since multiple paths exist for every OD pair, we consider the path combinations, $\mathit{C}$, each of which consists of one path for every OD pair. We utilize $\mathit{CNN_{i}}$ to represent the path combination $\mathit{CNN_i}$ and the output of $\mathit{CNN_i}$ means that the path combination should be chosen or not. Specifically, the input of a CNN is the traffic pattern and the remaining buffer size of MEO satellites and GEO satellites. The output is a vector composed of two binary elements, where $\mathit{(1,0)}$ represents the path combination should be chosen, otherwise $\mathit{(0,1)}$ means not. We utilize the online training method to train the considered CNNs~\cite{wcm2}. The main parameters of the utilized CNN are shown in Table~\ref{simu-para}.

Figs.~\ref{fig-result} give the results of our proposal and the conventional SP method in terms of network throughput and packet loss rate. In Fig.~\ref{per1}, we can find the throughput of our proposal linearly increases with the growth of the source nodes' number when the number of source nodes is less than 1,200. For the conventional SP strategy, the value of network throughput is nearly the same as that of our proposal and also linearly increases at the beginning. However, when the number of source nodes is above 800, the network throughput of conventional strategy increases gradually slowly and nearly keeps a constant value after the source nodes are more than 1,000. Therefore, we can find our proposal can achieve much higher throughput than the conventional strategy when the source nodes' number is more than 800. Fig.~\ref{per2} shows the packet loss rate of two networks. It can be easily found that the two strategies nearly have the same performance when the source nodes are less than 700. However, after that, the packet loss rate of conventional strategy increases very fast, which means that the network running conventional strategy is congested when there are more than 700 source nodes and the congestion becomes more seriously with the growth of source nodes. For the deep learning based routing strategy, the network can still transfer all the packets to destinations before the number of source nodes increases above 1,100. After that, our proposal also lose some packets. We can conclude that compared with the conventional strategy, our proposal can significantly improve the network performance. The reason for improvement of our proposal is that the utilized deep learning architecture can learn the experiences from traffic patterns and transfer some packets via GEO satellites when some MEO satellites are nearly congested. On the other hand, in the network running conventional strategy, the MEO satellites are always first chosen for packet transmission, which leads to frequent traffic congestion in the MEO satellites. 
 
\section{Future Directions}
\label{future-directions}
Deep learning is a promising technique to optimize the SAGIN performance since its efficiency and flexibility enable it to be a prospective paradigm to promote the SAGIN to a higher level. On the basis of current achievements, future research can be conducted in following four aspects. 

\textbf{\emph{Architecture Construction:}} When designing the deep learning based proposal, the first challenge is to construct the deep learning architecture. 
As there are so many deep learning architectures of which each can be utilized for different applications, it is still tough even for an experienced researcher to choose the suitable structure. Moreover, the input and output design of the architecture is deeply concerned with the studied problems and impacts on the prediction accuracy as well as the computation overhead. Considering more factors usually means not only much higher accuracy rate but also the increased training and running time. Furthermore, as the SAGIN is a multi-dimensional network which is more complex than current terrestrial networks, how to design the corresponding deep learning architectures still needs more consideration.

\balance 
\textbf{\emph{Proposal Deployment:}} As we mentioned above, the long latency of the space communication system is a serious shortcoming for the SAGIN. When we deploy deep learning based proposals, if we consider the centralized control manner, the computation time cost as well as the transmission latency between the controller and the controlled devices will deteriorate the network performance. However, the distributed control is not always a good choice for any scenario, especially when the input of deep learning architecture is concerned with the whole network. Therefore, the distributed control manner often means increased signaling overhead caused by necessary information exchange. Moreover, the device computation requirement in the distributed network is much higher because every device needs to run the deep learning based method.  

\textbf{\emph{Computation Efficiency:}} Since the deep learning technique is usually concerned with much more computation consumption than conventional methods, when considering the deep learning technique to optimize the SAGIN performance, it is necessary to improve the computation efficiency. The computation efficiency is concerned with many factors, such as the required accuracy rate, the training strategy, the computation hardware, and so on. In most cases, increasing the number of layers can improve the accuracy rate, but also leads to increased computation consumption. Therefore, when designing the deep learning architectures, we need to consider the accuracy requirement and try our best to simplify the architectures. Many other techniques to improve the computation efficiency should also be considered, such as utilizing the parameters of existing deep learning architectures for initialization. Furthermore, the offline training manner can reduce the computation overhead during running phase, while the online training fashion enables the deep learning architecture to easily adapt to the changing patterns. 

\textbf{\emph{Hardware Design:}} To deploy the deep learning based strategies, we also need to make some modifications on current communication hardware. It is common that the GPU is more suitable for deep learning computation since it can efficiently handle the matrix computations. Thus, if we adopt the deep learning technique to optimize the SAGIN performance, it is important to equip enough GPU resource in current network infrastructure. At the same time, the cooperation among the different computation resource as well as the computation task scheduling also need to be studied. Moreover, the hardware expense should be also considered to improve the cost performance ratio, especially for the space communication systems. 

\section{Conclusion}
\label{conclusion}
In this paper, we study the deep learning technique to optimize the SAGIN performance. Firstly, we discuss several challenges which directly impact on the network performance. Then, we give some analysis about existing deep learning based research to improve the terrestrial network performance and also explain the shortcomings of these proposals for future SAGIN. To show how the deep learning technique can improve the SAGIN performance, we give an example which utilizes the CNN to choose the path combinations for the satellite communication system. Since applying the AI technique in SAGIN is still a new topic, we also discuss some future promising directions.


\begin{thebibliography}{10}
	
	\bibitem{shen-5g}
	N.~Zhang, P.~Yang, J.~Ren, D.~Chen, L.~Yu, and X.~Shen, ``{Synergy of Big Data
		and 5G Wireless Networks: Opportunities, Approaches, and Challenges},'' {\em
		IEEE Wireless Communications}, vol.~25, pp.~12--18, Feb. 2018.
	
	\bibitem{satellite-survey}
	C.~Niephaus, M.~Kretschmer, and G.~Ghinea, ``{QoS Provisioning in Converged
		Satellite and Terrestrial Networks: A Survey of the State-of-the-Art},'' {\em
		IEEE Communications Surveys Tutorials}, vol.~18, pp.~2415--2441,
	Fourthquarter 2016.
	
	\bibitem{air-com}
	``{The Flight to Own the Stratosphere: Google's Project Loon vs Facebook's
		Aquila}.''
	\url{https://www.yaabot.com/29834/battling-it-out-for-internet-access-from-the-skies-googles-project-loon-versus-facebooks-aquila/
	}, (accessed Jun. 2018).
	
	\bibitem{liu-sagin}
	J.~Liu, Y.~Shi, Y.~Shi, Z.~M. Fadlullah, and N.~Kato, ``{Space-Air-Ground
		Integrated Network: A Survey},'' {\em IEEE Communications Surveys Tutorials},
	2018, doi: 10.1109/COMST.2018.2841996.
	
	\bibitem{zubair-survey}
	Z.~M. Fadlullah, F.~Tang, B.~Mao, N.~Kato, O.~Akashi, T.~Inoue, and
	K.~Mizutani, ``{State-of-the-Art Deep Learning: Evolving Machine Intelligence
		Toward Tomorrow's Intelligent Network Traffic Control Systems},'' {\em IEEE
		Communications Surveys Tutorials}, vol.~19, pp.~2432--2455, Fourthquarter
	2017.
	
	\bibitem{wcm1}
	N.~Kato, Z.~M. Fadlullah, B.~Mao, F.~Tang, O.~Akashi, T.~Inoue, and
	K.~Mizutani, ``{The Deep Learning Vision for Heterogeneous Network Traffic
		Control: Proposal, Challenges, and Future Perspective},'' {\em IEEE Wireless
		Communications}, vol.~24, pp.~146--153, Dec. 2017.
	
	\bibitem{mao2017routing}
	B.~Mao, Z.~M. Fadlullah, F.~Tang, N.~Kato, O.~Akashi, T.~Inoue, and
	K.~Mizutani, ``{Routing or Computing? The Paradigm Shift Towards Intelligent
		Computer Network Packet Transmission Based on Deep Learning},'' {\em IEEE
		Transactions on Computers}, vol.~66, pp.~1946--1960, Nov. 2017.
	
	\bibitem{l-band}
	T.~Pelzmann, T.~Jost, M.~Schwinzerl, F.~Pérez-Fontán, M.~Schönhuber, and
	N.~Floury, ``{Airborne Measurements Enhancing the Satellite-to-aircraft
		Channel Model in L-band},'' in {\em 2016 10th European Conference on Antennas
		and Propagation (EuCAP)}, pp.~1--5, Apr. 2016.
	
	\bibitem{energy-routing}
	Y.~Yang, M.~Xu, D.~Wang, and Y.~Wang, ``{Towards Energy-Efficient Routing in
		Satellite Networks},'' {\em IEEE Journal on Selected Areas in
		Communications}, vol.~34, pp.~3869--3886, Dec. 2016.
	
	\bibitem{EE-UAV}
	Y.~Zeng and R.~Zhang, ``{Energy-Efficient UAV Communication With Trajectory
		Optimization},'' {\em IEEE Transactions on Wireless Communications}, vol.~16,
	pp.~3747--3760, Jun. 2017.
	
	\bibitem{uav-survey}
	L.~Gupta, R.~Jain, and G.~Vaszkun, ``{Survey of Important Issues in UAV
		Communication Networks},'' {\em IEEE Communications Surveys Tutorials},
	vol.~18, pp.~1123--1152, Secondquarter 2016.
	
	\bibitem{tang-poc}
	F.~Tang, Z.~M. Fadlullah, B.~Mao, and N.~Kato, ``{An Intelligent Traffic Load
		Prediction Based Adaptive Channel Assignment Algorithm in SDN-IoT: A Deep
		Learning Approach},'' {\em IEEE Internet of Things Journal}, online, 2018.
	
	\bibitem{8290944}
	X.~He, K.~Wang, H.~Huang, T.~Miyazaki, Y.~Wang, and S.~Guo, ``{Green Resource
		Allocation based on Deep Reinforcement Learning in Content-Centric IoT},''
	{\em IEEE Transactions on Emerging Topics in Computing}, online, 2018.
	
	\bibitem{DBLP:journals/corr/abs-1710-00811}
	A.~Tuor, S.~Kaplan, B.~Hutchinson, N.~Nichols, and S.~Robinson, ``{Deep
		Learning for Unsupervised Insider Threat Detection in Structured
		Cybersecurity Data Streams},'' {\em CoRR}, vol.~abs/1710.00811, 2017.
	
	\bibitem{wcm2}
	F.~Tang, B.~Mao, Z.~M. Fadlullah, N.~Kato, O.~Akashi, T.~Inoue, and
	K.~Mizutani, ``{On Removing Routing Protocol from Future Wireless Networks: A
		Real-time Deep Learning Approach for Intelligent Traffic Control},'' {\em
		IEEE Wireless Communications}, vol.~25, pp.~154--160, Feb. 2018.
	
\end{thebibliography}


\end{document}